# Mining unit test cases to synthesize API usage examples

Mohammad Ghafari[1]  |  Konstantin Rubinov[2]  |  Mohammad Mehdi Pourhashem K.[2]

[1]Software Composition Group, University of Bern, Switzerland

[2]DeepSE Group, Politecnico di Milano, Italy

**Correspondence**
Mohammad Ghafari, Software Composition Group at University of Bern, Switzerland.
Email: ghafari@inf.unibe.ch

**Funding information**
Swiss National Science Foundation, Grant/Award Number: 200020-162352

**Abstract**

Software developers study and reuse existing source code to understand how to properly use application programming interfaces (APIs). However, manually finding sufficient and adequate code examples for a given API is a difficult and a time-consuming activity. Existing approaches to find or generate examples assume availability of a reasonable set of client code that uses the API. This assumption does not hold for newly released API libraries, non-widely used APIs, nor private ones.

In this work we reuse the important information that is naturally present in test code to circumvent the lack of usage examples for an API when other sources of client code are not available. We propose an approach for automatically identifying the most representative API uses within each unit test case. We then develop an approach to synthesize API usage examples by extracting relevant statements representing the usage of such APIs. We compare the output of a prototype implementation of our approach to both human-written examples and to a state-of-the-art approach. The obtained results are encouraging; the examples automatically generated with our approach are superior to the state-of-the-art approach and highly similar to the manually constructed examples.

## 1 | INTRODUCTION

Software reuse is a practice of using existing software, or software knowledge, to construct new software. Reusing existing software can decrease development effort while increasing the quality of production, if mature previously tested assets are used. Developers cope with the complexity of modern software systems and speed up the development process by increasingly relying on the functionalities provided by off-the-shelf components and frameworks. However, an effective integration of external third-party libraries requires an extensive knowledge of the application programming interfaces (APIs) and a detailed understanding of the required interaction protocols. Unfortunately, acquiring these competences is difficult and time-consuming. Moreover, real-world software libraries and frameworks may be underspecified or poorly documented. In addition, if documentation exists, it may include erroneous or out-of-date content.[1] Hence, software developers resort to studying code examples to learn APIs, and pragmatically reuse parts of a code corpus that exercise these APIs to reduce the development effort.[2]

Given these premises, code examples play a crucial role in modern software development complementing the existing documentation and facilitating the learning curve for developers. However, generating meaningful code examples for a given API and finding the relevant ones is a difficult and time-consuming activity. Robillard et al. show that, in practice, insufficient or inadequate examples represent the greatest obstacle to effectively learn an API.[2] This aspect becomes even more relevant during software maintenance where maintainers usually have to deal with multiple different APIs that they may have exercised neither recently nor frequently.

In response to the increasing need of meaningful code examples, research has been focusing on techniques to automatically extract or generate them to assist developers.[3,4] To find or generate the examples, existing approaches typically rely on external sources such as the source code of existing local or remote projects.[5–7] More recently, good coverage of API usage examples in Q&A websites motivated researchers to use these crowd sources as well, for both documentation[8] and recommendation purposes.[9,10] Regardless of whether developers explore these resources manually or use a recommender system, there are still several problems that hamper the applicability of existing approaches in practice.[11] Among them, untrustworthiness, low quality, and inaccessibility of appropriate code examples are the most important ones. Also, maintaining these resources is challenging partly because they are not explicitly linked to the APIs, and changes in the APIs are not reflected in the resources.[1] Furthermore, although existing approaches excel at locating examples of frequent APIs, they all suffer from the "cold-start" problem, i.e., examples can only be extracted when a reasonable sample of client systems that use the API exist.[12] However, this is a big assumption. These approaches fall short at proposing examples when client program is not available (e.g., in case of private APIs) or clients themselves do not exist yet (e.g., for newly released API libraries, or nonwidely used ones).

Agile development is founded on a seamless integration between continuous code changes and unit testing. Unit test cases are small fragments of code in charge of testing a unit of work in the system by checking a single assumption about the behavior of that unit of work. Once developers have written the test cases, they are executed every time the production code changes to support regression testing. This process requires unit test cases to be always up-to-date and makes them an important source of system documentation, especially for guiding software maintenance tasks. Unit test cases usually cover at least the critical functionalities provided by an API. Studying unit test cases of an API conveys significant information on (1) how to correctly instantiate classes, (2) how to construct arguments for method calls, and (3) the expected system state after the invocations of methods. Furthermore, unit test cases are not only helpful for API users but also advantageous to API developers. In software development, not all the APIs are designed to be exposed to end users: some private APIs are intended to help the API developers during development process. Such APIs are internal, and therefore, there is no usage example for them in external resources. Also, they often lack usage documentation because of constraints in the development schedule. Hence, unit test cases of private APIs could be a good resource of such API uses. The amount of code that exists in test suites is limited in size, and at the same time, it is highly relevant with respect to the tested API. Unit test cases are usually executable, simple, and concise snippets that are conceived to run in isolation.

In the light of these considerations, test cases seem to be a significant source of API uses in the absence of client systems.[11,13] However, our experiments with state-of-the-art tools on several open-source libraries showed that even if test cases are part of the repository of these tools, they cannot be directly used to provide effective results. In fact, the different structure of test code with respect to general code suggests adoption of a dedicated mining approach to obtain examples from test cases.

This paper presents an approach for automatically extracting API usage examples from test cases written in JUnit. The proposed approach intends to complement approaches that extract API usage examples from source code of client systems and is intended to be used when these client systems are not (yet) available and the only source of such examples is the test code. The approach is general; we detail each step of the approach with some reference examples taken from different systems. The approach consists of 2 phases: in the first phase it identifies traceability links between test methods and focal methods under test (F-MUTs). Our previous work has shown that focal methods represent the core of a test scenario inside a unit test case, and each test case represents more useful information about the F-MUTs rather than other (non-focal) methods.[14] In the second phase we bring our work one step further, extend it by developing a relevant use case—synthesizing API usage examples from test code—and practically show that the F-MUT information is advantageous to accomplish this task.

The approach traces object state changes verified in the oracle part of a test case to identify F-MUTs. To construct examples, the approach examines the type of dependency relation between F-MUTs to determine if each of them should contribute to a different example. We adapted a graph-based object usage model that superimposes control dependency graph and data dependency graph to illustrate how each focal method interacts with other code elements within a test. The model takes into account the test semantics and the role of different methods within a test case. The approach uses this model to identify necessary statements for representing the usage of each F-MUT within a test. The proposed approach has been implemented in a prototype, and both phases of the approach have been experimentally evaluated on different open-source systems. We have compared the examples generated with our approach to both human-written examples and examples generated by UsETeC—a state-of-the-art tool of Zhu et al.[4] The results show that our approach is superior to the state-of-the-art approach and generates examples that are highly similar to the manually constructed examples.

In summary, this work makes the following contributions:

- a formal representation of an approach to recover test-to-code traceability links on the method level precision by identifying the most representative APIs (focal methods) in unit test cases;
- an approach to automatically synthesize meaningful usage examples of the focal methods from unit test cases;
- a prototype implementation of the example synthesis approach in Java, and the evaluation of its effectiveness on 4 different real-world software systems.

The remainder of this paper is organized as follows. Section 2 presents our automated approach to identifying F-MUTs, and Section 3 synthesizes meaningful API uses w.r.t. each focal method within a unit test case. Section 4 evaluates the prototype implementation of the proposed approach on real-life software systems and discusses the results. Section 5 overviews related work, and Section 6 concludes the paper.

## 2 | IDENTIFYING TRACEABILITY LINKS BETWEEN SOURCE AND TEST CODE

To enable automated extraction of relevant API usage information from test code, we must first establish the relationship between the source and test code to be analyzed. Previous research has mostly focused on deriving this relationship only between test cases and classes under test.[15] Although the knowledge of the class under test (CUT) is useful for test case comprehension and analysis, CUT information is insufficient to analyze test cases with a method-level precision. Our previous research has shown that a unit test case represents more useful usage information of the focal method rather than other (non-focal) methods within that test case, which led to the development of a solution for their automatic identification.[14] In the rest of this section we introduce the most relevant concepts and background in identifying F-MUTs.

```
1  public void testRegisterAndRemoveProxy() {
2    // register a proxy, remove it, then try to retrieve it
3    IModel model = Model.getInstance();
4    IProxy proxy = new Proxy("sizes", new String[]{"7","13","21"});
5    model.registerProxy(proxy);
6    // remove the proxy
7    IProxy removedProxy = model.removeProxy("sizes");
8    // assert that we removed the appropriate proxy
9    assertEquals(removedProxy.getProxyName() , "sizes");
10   // ensure that the proxy is no longer retrievable from the model
11   proxy = model.retrieveProxy("sizes");
12   assertNull("Expecting proxy is null", proxy);
13 }
```

**FIGURE 1** Unit test case for the `Model` class

```
1  public void testEraseAll() {
2    ArrayTable<String, Integer, Character> table
3      = create("foo", 1, 'a', "bar", 1, 'b', "foo", 3, 'c');
4    table.eraseAll();
5    assertEquals(9, table.size());
6    assertNull(table.get("bar", 1));
7    assertTrue(table.containsRow("foo"));
8    assertFalse(table.containsValue('a'));
9  }
```

**FIGURE 2** Unit test case for the `ArrayTable` class

## 2.1 | Test case structure

The core ideas behind our approach are formed from a few observations of a large number of real-life open-source projects. We observed test cases of different types and granularity use method invocations as the atoms to construct test cases from the work of Pezze et al.[16] Our analysis indicates that test cases invoke on average 6 methods. We have manually investigated the role of each method within test cases with more than 1 method invocation. The result shows most of these methods are ancillary to few that are intended to be the actual (or **focal**) methods under test. Focal methods represent the core of a test scenario inside a test case, and therefore, each test case represents more useful information about the **F-MUT** rather than other (non-focal) methods. Consequently, identifying F-MUTs is a preparatory step for synthesizing useful API usage examples from test code.

Consider an example unit test case from *PureMVC*[*] in Figure 1. In the example, `IModel` is the CUT. Three methods in the test case belong to the CUT `IModel` any of which can be the method under test. However, the real intent in this test case is to check the `removeProxy()` method. An expert engineer can identify this with the aid of comments, test method names, and assertions. Without this knowledge or additional analysis, one might mistakenly conclude that the goal of the test is to check `registerProxy()` or `retrieveProxy()` methods. Although method `registerProxy()` might be viewed as a relevant method to the test case, its role is ancillary; it brings the `model` object to an appropriate state in which it is possible to invoke the `removeProxy()` method. The `retrieveProxy()` invocation is used to inspect the state of the CUT while it is the method `removeProxy()` that causes a side effect on the current object and is the focal method under test.

Unit test cases are commonly structured in 3 logical parts: *setup*, *execution*, and *oracle*. The *setup* part instantiates the CUT and includes all dependencies on other objects that the unit under test will use. This part contains initial method invocations that bring the object under test into a state required for testing. The *execution* part stimulates the object under test via a method invocation, i.e., the focal method in the test case. This action is then checked with a series of inspector methods and `assert` statements in the *oracle* part that query the side effects of the focal invocation to determine whether the expected outcome is obtained.

In the context of object-oriented systems, unit test cases often test a single method.[17] Nevertheless, occasionally, test cases aggregate and test several methods in a test scenario. In this case a complete *test scenario* comprises several sub-scenarios, where a *sub-scenario* contains a set of non-assert statements (setup and execution) followed by inspector and `assert` statements. That is, each sub-scenario may have a different focal method, and therefore, a test case can have more than 1 focal method.

A focal method belongs to the execution part of a test case, and method invocations used in the oracle part often only inspect the side effect of the F-MUT. Despite the clear logical differentiation of test parts, each having its own purpose, in practice the parts are hardly discernible either manually or automatically. This hinders identifying F-MUTs without expert knowledge of the system. It is difficult to establish whether a method invocation belongs to the setup or the execution parts of a test. Even the oracle part associated with `assert` statements may contain method invocations that may be confused with the execution part of the test case.

In practice, data flow analysis is required to distinguish among different types of method invocations in test cases. We need to distinguish F-MUTs from inspector methods serving the oracle part. For example, the test case in Figure 2 belongs to the `ArrayTableTest` class in *Guava* library.[†] There are 6 method invocations within this test case. The first invocation `create()` belongs to the setup part of the test case. This invocation is a helper method that initializes the `table` object and puts this object in an appropriate state for testing. The invocation at line 4, `table.eraseAll()` is the F-MUT and belongs to the execution part of the test case. In fact, the method `eraseAll()` causes a state change of the `table` object, whose effects are later inspected using 4 other invocations, namely, `table.size()`, `table.get()`, `table.containsRow()`, and `table.containsValue()`. Invocations at lines 5 to 8 are *inspector methods* and contribute to the oracle part of the test by inspecting the state of the object under test affected by the focal method `eraseAll()`, while preserving values of the class fields.

---
[*] http://puremvc.org

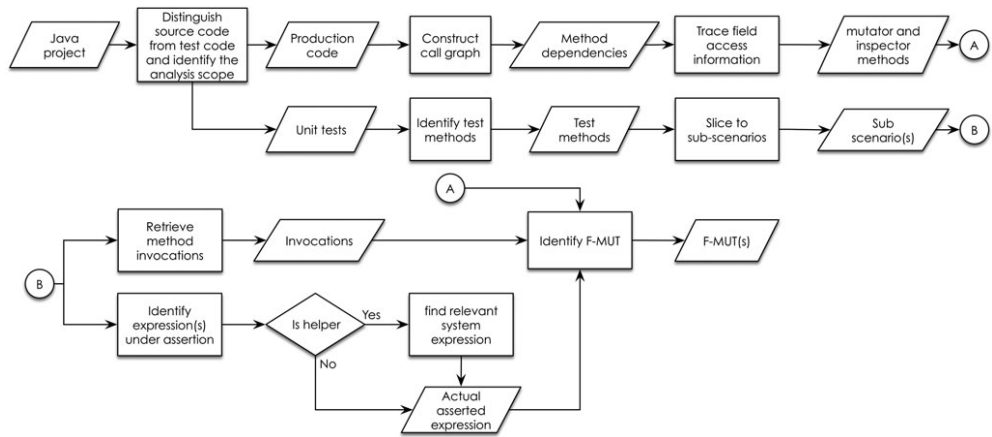

**FIGURE 3** The main steps of the proposed approach for identifying focal methods under test (F-MUTs)

## 2.2 | Identifying focal methods under test

Having observed that identification of F-MUTs is not trivial and requires custom analysis, we have developed an approach on the basis of a novel heuristic to support the developer in the identification of the F-MUTs in a unit test case.[14] By investigating a large number of open-source programs, we observed that dependencies between F-MUTs, the CUT and assertions manifest themselves through object state changes verified in the oracle part of a test case. This observation forms the underlying intuition behind our approach. Furthermore, our empirical observations led to the following heuristic:

> The last method invocation entailing an object state change whose effect is inspected in the oracle part of a test case is a focal method under test (F-MUT).

Our approach leverages data flow analysis to capture essential information in test cases and source code. It is general and applies to object-oriented systems. In this work we instantiate the approach for projects in Java and test cases in JUNIT format, the de-facto standard for unit testing Java applications.[‡] Our static analysis works on the abstract syntax tree representation of the source code. The steps of the approach and the input/output for each step are shown in Figure 3. In the following we provide an overview of these steps, but the interested reader may consult our previous work for a thorough discussion.[14] To explain our heuristic approach, we introduce a simple set-theoretic model in Table 1 that captures the key elements of the system. We also present the formal definitions of the essential notions used in the approach in Table 2.

The approach takes as input a Java project and extracts test cases from the code being tested. It then analyzes test cases to establish the scope of the analysis, i.e., which system classes are involved in testing. It analyzes the identified classes of the system *C* (source code) to extract system dependencies and construct a system call graph *CallGraph* $\subseteq 2^{M \times M}$ that represents the calling relationship between the methods of all the classes involved in testing (*Constructing call graph* in Figure 3).

A *CallGraph* is a directed graph whose nodes are methods in *M*, and each node $m_i$ is connected to $m_j$, iff $m_i$ invokes $m_j$. *CallChain*:$M \mapsto 2^M$ is a function that returns all the methods invoked directly/indirectly by the given method. In fact, *CallChain* of a given method is the set of methods reachable from that method in the *CallGraph*.

$$CallChain(m_1) = \{m_2 | (m_1, m_2) \in CallGraph \text{ or } \exists m' \text{ s.t. } m' \in CallChain(m_1), (m', m_2) \in CallGraph\}$$

The approach applies inter-procedural forward reachability analysis to detect mutator and inspector methods within the classes of interest. An inspector is a side-effect free method that returns information about the state of an object, whereas a method causing object state change is a mutator method. To determine whether a method is mutator or inspector, our approach analyzes the object fields in which a method accesses in its method body and also keeps track of changes to method parameters (*tracing field access information* in Figure 3).

We suppose an *action* is the most fine-grained operation on an object including its instantiation, invocation, or field access. Accordingly, *Action* represents the set of all actions in the system. $M_{Actions}$:$M \mapsto 2^{Action}$ is a function that returns the set of actions within the body of a given method, and $M_{Actions^*} : M \mapsto 2^{Action}$ is a function that beside such actions ($M_{Actions}$) also returns all the actions in the body of all other methods

---
[†]http://code.google.com/p/guava-libraries
[‡]http://junit.org

**TABLE 1** The core model of the system

| Formal Element | Description |
|---|---|
| $C$ | The set of all the classes. |
| $M$ | The set of all the methods. |
| $V$ | The set of all the variables. |
| Action | The set of all actions in the system. |
| Assertion | The set of all the assertion statements. |
| $Acc = \{r, w\}$ | The set of **r**ead/**w**rite access types on variables. |
| $Act \subseteq Action \times V \times Acc$ | The set of actions' activities on the variables. |
| $C_M: C \mapsto 2^M$ | Receives a class and returns a set of methods that belong to the given class. |
| $C_V: C \mapsto 2^V$ | Receives a class and returns a set of variables that belong to the given class. |
| $M_V: M \mapsto 2^V$ | Receives a method and returns its arguments. |
| $M_{Actions}: M \mapsto 2^{Action}$ | Receives a method and returns its actions. |
| $M_{Actions}^*: M \mapsto 2^{Action}$ | Receives a method and returns its actions as well as the actions belonging to its CallChain. |

**TABLE 2** The formal notions

| Notion | Definition |
|---|---|
| CallGraph | A directed graph whose nodes are methods in $M$ and each node $m_i$ is connected to $m_j$, iff $m_i$ invokes $m_j$. |
| CallChain | A function that returns all the methods invoked directly or indirectly by a given method. CallChain$(m_1) = \{m_2 \mid (m_1, m_2) \in$ CallGraph or $\exists m'$ s.t. $m' \in$ CallChain$(m_1), (m', m_2) \in$ CallGraph$\}$ |
| Mut | The set of mutators that comprises pairs of a method and a mutated variable, $(m,v)$. Mut $= \{(m, v) \mid \exists a_i \in M_{Actions}^*(m), c \in C$ s.t. $m \in C_M(c), v \in (C_V(c) \cup M_V(m)), (a_i, v, w) \in Act\}$ |
| Ins | The set of inspectors that comprises pairs of a method and an inspected variable, $(m,v)$. Ins $= \{(m, v) \mid \exists a_i \in M_{Actions}^*(m), c \in C$ s.t. $m \in C_M(c), v \in (C_V(c) \cup M_V(m)), (a_i, v, w) \in Act$ and $(m, v) \notin$ Mut$\}$ |
| TM | A test method is a set of sub-scenarios (ss), each of which is a sequence of actions followed by a sequence of assertion statements. TM $= \{a_i \ldots a_n t_j \ldots t_m \mid \{a_i, \ldots, a_n\} \subseteq$ Action and $\{t_j, \ldots, t_m\} \subseteq$ Assertion, $n \geqslant i, m \geqslant j$ and $i, j > 0\}$ |
| $FM_{ss}$ | Focal Method of a sub-scenario is the last mutator having side effect on the actual asserted expression. $FM_{ss}(ss) = \{m_1 \mid \exists v \in V, a_i \in$ Action, $t_k$ s.t. $m_1$ is invovled in $a_i$ and $v$ is asserted in $t_k, (m_1, v) \in$ Mut and $\nexists j > i$ s.t. $m_2 \in$ Method, $m_2$ is involved in $a_j \in$ Action and $(m2, v) \in$ Mut$\}$ |
| $FM_{tm}$ | A function that returns the focal methods of a given test method. $FM_{tm}(tm) = \cup_{i=1}^n FM_{ss}(ss_i)$ |

reachable from the given method. This function, called *method chain actions*, is defined as follows:

$$M_{Actions}^*(m) = \{a \mid a \in M_{Actions}(m) \text{ or } \exists m' \text{ s.t. } a \in M_{Actions}(m') \text{ and } m' \in CallChain(m)\}$$

Action $a$ is a member of the returned set of $M_{Actions}^*(m)$, iff $a$ belongs either to $m$ or to one of the methods in its *CallChain*.

$Mut \subseteq M \times V$ is the set of mutators that comprises pairs of a method and a mutated variable, $(m,v)$. Method $m$ is a mutator of variable $v$, iff there exists an action in the method chain actions of $m$, such that it has a write access on $v$, which is either a parameter of $m$ or a member of $m$'s class.

$$Mut = \{(m, v) \mid \exists a_i \in M_{Actions}^*(m), c \in C \text{ s.t. } m \in C_M(c),$$
$$v \in (C_V(c) \cup M_V(m)), (a_i, v, w) \in Act\}$$

Where $C_M$ and $C_V$ are the sets of methods and variables of the given class, respectively; and $M_V$ is the set of arguments of the given method.

$Ins \subseteq M \times V$ is the set of inspectors that comprises pairs of a method and an inspected variable, $(m, v)$. Method $m$ is an inspector of variable $v$, iff there exists an action in the method chain actions of $m$, such that it has only read access on $v$ that is either a parameter of $m$ or a member of $m$'s class.

$$Ins = \{(m, v) \mid \exists a_i \in M_{Actions}^*(m), c \in C \text{ s.t. } m \in C_M(c),$$
$$v \in (C_V(c) \cup M_V(m)), (a_i, v, w) \in Act \text{ and}$$
$$(m, v) \notin Mut\}$$

Afterwards, the approach follows the test naming convention in JUnit 3 and annotations in JUnit 4 to distinguish test methods *TM* from helper ones (*identifying test methods* in Figure 3). It partitions each test method into sub-scenarios (ss), each of which is a sequence of actions followed by a sequence of assertion statements (*slicing to sub-scenarios* in Figure 3).

$$TM = \{a_i \ldots a_n t_j \ldots t_m | \{a_i, \ldots, a_n\} \subseteq \textit{Action} \text{ and } \{t_j, \ldots, t_m\} \subseteq \textit{Assertion},$$
$$n \geqslant i, m \geqslant j \text{ and } i, j > 0\}$$

We consider all the common overloaded variants of the `assert` statements in JUnit format. If an asserted expression is a single variable, we find the method invocation from which the variable is assigned in that unit test case. If the declaring class of this variable does not originate in the project source code, we mark it as a helper[§] (*identifying expression(s) under assertion* in Figure 3). The F-MUTs affect the state of the classes of the system under test, and assertions may check these state changes indirectly by accessing helper variables, rather than directly by accessing the classes of the system under test. In this case, we search, within a test method, for an expression from which the helper class is instantiated. The search continues recursively until we find an invocation or a field access on a class of the system under test that instantiates the helper. We register this invocation as the actual asserted expression (*finding relevant (asserted) system expression* in Figure 3).

Having a set of actual asserted expressions for each sub-scenario *ss*, and the knowledge of mutator and inspector methods that our analysis discovered from the system source code, the approach reports a focal method $FM_{ss}: TM \mapsto 2^M$ that is a last mutator having a side effect on the actual asserted expression in that sub-scenario (*identifying F-MUT* in Figure 3).

$$FM_{ss}(ss) = \{m_1 | \exists v \in V, a_i \in \textit{Action},$$
$$t_k \text{ s.t. } m_1 \text{ is invovled in } a_i \text{ and } v \text{ is asserted in } t_k,$$
$$(m_1, v) \in \textit{Mut} \text{ and } \nexists j > i \text{ s.t. } m_2 \in \textit{Method},$$
$$m_2 \text{ is involved in } a_j \in \textit{Action and}$$
$$(m_2, v) \in \textit{Mut}\}$$

Finally, $FM_{tm}: 2^{TM} \mapsto 2^M$ is a function that returns the focal methods of the given *TM*, that is, essentially,

$$FM_{tm}(tm) = \cup_{i=1}^{n} FM_{ss}(ss_i), \text{ where } tm = \{ss_1, ss_2 \ldots, ss_n\}$$

The approach suffers from several limitations inherent to static analysis approaches that are generally unsound. For instance, the call graph construction bears limitations of the approach on type resolution for interface calls and polymorphic method calls. Moreover, we apply a simplified intra-procedural alias analysis to identify references to a class field from a local variable. That is, when tracing field access information, possible state manipulations may escape when a method parameter references a class field. We have thoroughly discussed these limitations in our dedicated work to F-MUT identification and have shown the significant effectiveness of the approach to identify F-MUTs for real-world software systems, despite the identified limitations.[14]

## 3 | SYNTHESIZING API USES

In the first phase of the approach (Section 2) we identified F-MUTs that are the most representative uses of an API within a unit test case. This section presents the second phase of the approach that automatically synthesizes a set of examples representing the usage of such APIs.

Not all sequences of API calls in test code represent meaningful usage examples. In fact, a test case can have more than 1 focal method, and each focal method may focus on a different aspect of the unit under test that represents a separate API use. Figure 4 shows a real test case from the PureMVC project. The scenario in the test case is as follows. First, a `controller` object is created. Then, `controller`'s `registerCommand` is called to register a particular `ICommand` class, `ControllerTestCommand`, as the handler for a particular `INotification`, named `hasCommandTest`. Next, the result of `controller.hasCommand` is checked through a JUnit method `Assert.assertTrue` to determine if the command is successfully registered for the specified notification. In the second sub-scenario, invocation of the focal method `controller.removeCommand` removes the previously registered command checked with the `Assert.assertFalse` JUnit method. The sequence of API uses `registerCommand` and then immediately `removeCommand` only serves a testing purpose; it is not a usage scenario that solves a practical programming problem. More specifically, each of these focal methods represents a **separate** example.

Figure 5 shows another unit test case from the PureMVC project comprising 2 sub-scenarios for exercising `registerCommand` and `executeCommand` methods. The first sub-scenario is similar to the one explained for the unit test case in Figure 4. In the second sub-scenario, a helper object named `vo` is created and used as an optional body in the creation of the `note` object in line 7. Afterwards, invocation of the `executeCommand` on the `controller` object executes the `ICommand` previously registered as the handler for the given notification at line 3. This call is expected to change the value of the `result` attribute in the `vo` object. This is determined at the last line. In contrast to the previous test case where each focal method represented a separate API usage example, the focal methods in the current test are relevant and **together** represent a useful example.

Therefore, the temporal sequence of API uses (focal methods) in test code is not always similar to the intended API uses as they appear in client code. Determining whether a sequence of 2 focal methods should be separated or not is not possible without realizing the relation between these methods. Moreover, to ease developer's understanding, usage examples should be concise and free from extraneous statements. That is, the mined examples must be processed before being presented to the user, to eliminate statements that are not necessary, like of those only belonging to the oracle part, e.g., inspector methods and assertion statements. In the rest of this paper we use the term **meaningful example** in referring to a code snippet that represents such an API use.

Finally, extracting meaningful examples from a given test may not be possible with traditional slicing proposals.[18] These techniques fall short of precisely identifying statements related to a particular focal method, as focal methods are typically involved with the same program entities. For instance, focal methods are usually invoked on the same object, i.e., the unit under test. Moreover, different F-MUTs within a unit test may use the same setup part. Due to this intrinsic shared data among focal methods in a unit test case, existing slicing techniques may intermingle different uses together, rather than constructing

---
[§]Commonly these instances belong to classes external to the system, e.g., libraries, mock objects and stubs.

```
1  public void testHasCommand() {
2    IController controller = Controller.getInstance();
3    controller.registerCommand("hasCommandTest", new ControllerTestCommand());
4    assertTrue(controller.hasCommand("hasCommandTest"));
5    // MetaComment: second sub-scenario starts here
6    controller.removeCommand("hasCommandTest");
7    assertFalse(controller.hasCommand("hasCommandTest"));
8  }
```

**FIGURE 4** Two sub-scenarios with separate API uses in 1 unit test case

```
1  public void testRegisterAndExecuteCommand() {
2    IController controller = Controller.getInstance();
3    controller.registerCommand("ControllerTest", new ControllerTestCommand());
4    if (controller.hasCommand("ControllerTest") == false)
5      Assert.fail("The specified notification has no command!");
6    // MetaComment: second sub-scenario starts here
7    ControllerTestVO vo = new ControllerTestVO(12);
8    Notification note = new Notification("ControllerTest", vo, null);
9    controller.executeCommand(note);
10   Assert.assertTrue("Expecting vo.result == 24", vo.result == 24);
11 }
```

**FIGURE 5** Two sub-scenarios with relevant API uses in 1 unit test case

```
1  public void testSetMethods() throws Throwable {
2    MockPartial mock = new MockPartial();
3    assertEquals(1970, mock.getYear());
4    assertEquals(1, mock.getMonthOfYear());
5    mock.setYear(2004);
6    assertEquals(2004, mock.getYear());
7    assertEquals(1, mock.getMonthOfYear());
8    mock.setMonthOfYear(6);
9    assertEquals(2004, mock.getYear());
10   assertEquals(6, mock.getMonthOfYear());
11   mock.set(2005, 5);
12   assertEquals(2005, mock.getYear());
13   assertEquals(5, mock.getMonthOfYear());
14   /* MetaComment: the rest of code is omitted to save space */
15 }
```

**FIGURE 6** Unit test case for the `BasePartial` class

a slice, which focuses attention on the API of interest (excluding the code irrelevant to the F-MUT execution). For instance, consider the unit test case from JodaTime[¶] project in Figure 6. Three focal methods, each of which represents a different API use, comprise the test scenario in this unit test case, where the 3 API uses are highlighted with rectangles in the code. Suppose we are interested to extract the statements relevant to the method `mock.set` at line 11. If we construct a backward slice on the `mock` at this line, the slice will contain all statements affecting this object too (lines 2-11). Indeed, since the `mock` object is shared in different sub-scenarios, the slice becomes larger than expected rather than only focusing on the statements affecting on the `mock.set` method.

### 3.1 | Identifying meaningful sequence of API uses

Unit test cases generally exercise sequences of method calls that create and mutate objects. They check the return value and the state of the object on which the method was invoked to determine if F-MUTs behave as expected. According to our observations, a dependency relation exists in 2 different granularity levels between the statements within sub-scenarios in a test case.

Figure 7 illustrates a real unit test case from the Ant[∥] project. As indicated by the rectangles in the code, the second sub-scenario at line 8 uses the File object `reportFile` constructed in the first sub-scenario at line 3. Thus, a successful construction of the URL object in the second sub-scenario depends on the File object in the first scenario. Consequently, an example corresponding to the F-MUT in second sub-scenario, i.e., `URL.openStream()`, should comprise relevant statements from the first sub-scenario too.

Nevertheless, when F-MUTs operate on the same object (e.g., see Figure 6), often a dependency relation with a finer granularity than the object level exists among API methods. Indeed, in unit testing, methods that access (read or write) the same object are usually tested together,[19,20] although these methods may not necessarily be logically dependent or used together. That is, the sequence of such methods in a

---
[¶]http://www.joda.org/joda-time

[∥]http://ant.apache.org

```
1  public void testNoFrames() throws Exception {
2    executeTarget("testNoFrames");
3    File reportFile = new File(System.getProperty("root"), "src/etc/testcases/
       taskdefs/optional/junitreport/test/junit-noframes.html");
4    assertTrue("junit-noframes.html not present.", reportFile.exists());
5    assertTrue("Cant read the report file.", reportFile.canRead() );
6    assertTrue("File shouldnt be empty.", reportFile.length() > 0 );
7    // MetaComment: second sub-scenario starts here
8    URL reportUrl = new URL(FileUtils.getFileUtils().toURI(reportFile.
       getAbsolutePath()));
9    InputStream reportStream = reportUrl.openStream();
10   assertTrue("Stream shouldnt be empty.", reportStream.available() > 0);
11 }
```

**FIGURE 7**  Two explicitly data-dependent sub-scenarios in a unit test case

test may not be practical when used in the client code and thus hardly represents a meaningful usage example. Accordingly, we examine the type of dependency relation between the methods not only at object level on which they operate but also transitively at the level of object attributes to investigate whether a sequence of 2 F-MUTs (in short, sequence) is meaningful. Sequences can be of 3 kinds:

1. Read-Read. When 2 methods `m1` and `m2` both only read the same field `f`, each method can be executed separately without requiring another one. These methods are usually tested together as they need the same setup part.
2. Write-Read. If method `m1` writes field `f` and method `m2` reads the value written to field `f` by m1, the execution of `m2` depends on the value produced by `m1`.
3. Write-Write. When 2 methods `m1` and `m2` both write the same field `f`, and there is no Write-Read relation in between, they are independent and can be executed separately as one execution overwrites the other. Such sequences are very common in unit testing. For instance, many test scenarios involve testing the same method with different inputs.

The only sequence representing a meaningful usage scenario is Write-Read. Otherwise, each scenario can execute separately and represents a usage example.

### 3.2 | Computing relevant statements to an API

The F-MUTs within a test are syntactically dependent on one another and extracting the statements relevant to a focal method can be challenging. To realize how each focal method interacts with other code elements within a test, we adapt a graph-based object usage model that is more compact and specialized for object usage representation than program dependence graph and control flow graph.[21] We model the usage of each object including its instantiation, invocation, or field access—collectively called *actions*. The execution flow, data, and control dependencies among all of such actions form the entire *unit test model*.

In our model, multiple object usages in a test scenario can be represented with a directed acyclic graph. For instance, Figure 8 illustrates multiple objects participating in the unit test case in Figure 5. Nodes are labeled `C.m`, in which `C` is the class name of the object and `m` is the name of a method, constructor, or a field. The directed edges between actions represent the usage order, control or data dependency among them. We use different edge types to distinguish between the edges (see the legend in Figure 8). A directed edge from an action node `i` to an action node `j` represents the *temporal order* of actions w.r.t. the execution flow and indicates that in the test scenario, `i` is used before `j`, and `j` is used after `i`, respectively.

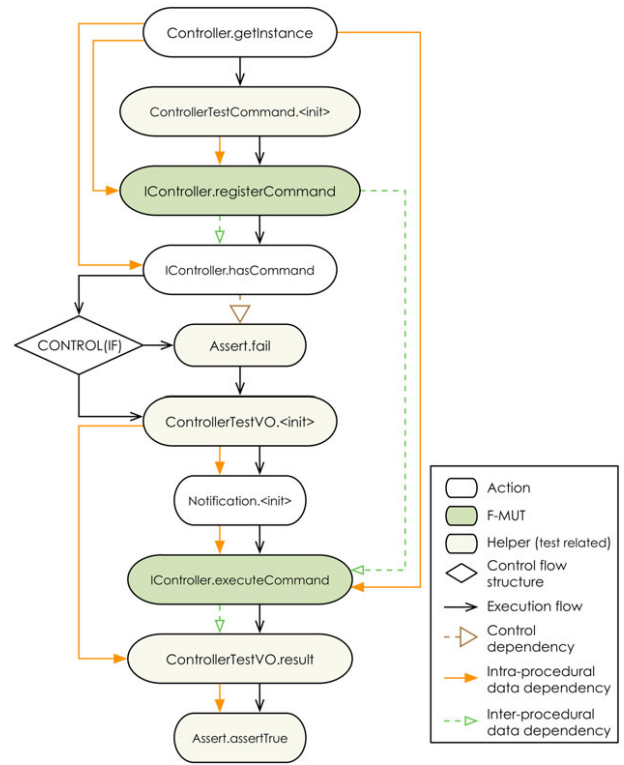

**FIGURE 8**  The representation of different object usages in the unit test case shown in Figure 5

Action $a_j$ is *control dependent* on action $a_i$ (control dependency CD between $a_i$ and $a_j$ evaluates to `true`) iff the execution result of $a_i$ determines whether $a_j$ will execute or not. Control flow CF: Action × Action ↦ {true, false} between $a_i$ and $a_j$ is the sufficient condition ($a_i.c_j$) for the former action to be followed by the latter action in a program execution: $\forall a_i \in$ Action if $Succ(a_i) = \{a_j, \ldots, a_n\}$, then $CF(a_i, a_j) = a_i.c_j, \ldots, CF(a_i, a_n) = a_i.c_n$, and $(\vee_{l=j}^{n} a_i.c_l) = $ true. In other words, (1) $CF(a_i, a_j) = $ false if $a_j$ is not a successor of $a_i$ , (2) $CF(a_i, a_j) = $ true if $a_j$ is the sole successor of $a_i$, and (3) if $a_i$ has more than one successor (there is a branch), then the disjunction of all CF from $a_i$ to its successors is true. For example, assume if *flag* is *true*, $a_1$ is followed by $a_2$, otherwise, it is followed by $a_3$. In such a branch, $CF(a_1, a_2) = a_1.c_2 = $ *flag*,

$CF(a_1, a_3) = a_1.c_3 = \neg flag$, and the disjunction of the 2 control flows is *true* so that one path will be chosen during the execution.

Algorithm 1 presents the control dependency computation on the basis of the conditions $a_i.Reachability$ on which each action $a_i$ is reachable from the root action $a_{root}$ in the control flow graph. The algorithm initializes the *WorkList* with the successors of the root (line 3). For each $a_i$ in the *WorkList* that is not yet processed (line 4), $a_i.Reachability$ expresses that $a_i$ is reachable, i.e., at least one of its predecessors $a_p$ is both reachable, and there is a control flow from $a_p$ to $a_i$. Once the algorithm computes the reachability conditions for an action (line 5), it performs the same computation for that action's successors; such computations continue until no unprocessed actions are left.

---

**Algorithm 1** Control dependency computation
1: **procedure** COMPUTECD()
2:     *Processed* ← {}
3:     *WorkList* ← *Succ(root)*
4:     **for** all $a_i \in$ *WorkList* \ *Processed* **do**
5:         $a_i.Reachability = \bigvee_{a_p \in Pred(a_i)} (a_p.Reachability \wedge CF(a_p, a_i))$
6:         *WorkList* ← *WorkList* ∪ *Succ*($a_i$)
7:         *Processed* ← *Processed* ∪ $a_i$
8:     **end for**
9: **end procedure**

---

In practice, we model control dependencies w.r.t. control statements such as `if` or `while` statements. We use control nodes to represent how actions are used within a control statement. To conform to the use of edges for representing temporal orders, we place control nodes at the branching node, where the program chooses an execution path, instead of placing them at the starting points of the corresponding statements. For example, the control node labeled CONTROL(IF) in Figure 8 represents the `if` statement in the test code in Figure 5, and the edge from the node IController.hasCommand to CONTROL(IF) indicates that the execution of hasCommand occurs before the branching point of the `if` control statement. All actions inside the body of a control structure are control dependent on the control predicate. The directed edge from the node IController.hasCommand to Assert.fail indicates a control dependency between these 2 actions in the test.

Action j is *data dependent* on action i, if i assigns to a memory location that j will read in some program path. In this work, we identify 2 kinds of data dependencies.

1. Intra-procedural. The object instance on which a particular method is invoked is data dependent on the object creation statement. For instance, there is a data dependency between Controller.getInstance and IController.registerCommand as the latter involves the object controller produced by the former action. Moreover, when an action is used as an argument in another action, e.g., ControllerTestCommand() and IController.registerCommand, the latter action is data dependent on the former.

2. Inter-procedural. In contrast to the previous data dependency, which we could spot by looking at each individual test in isolation, there may also be an inter-procedural dependency between 2 method calls that is not explicit from within a test. More specifically, to enable slicing test code w.r.t. a focal method, the precision we consider during dependency analysis should not be limited to an object level but we also trace transitive access on the object in different attributes to determine if 2 API methods are dependent. The dashed-edge between IController.registerCommand and IController.executeCommand in Figure 8 indicates such a data dependency between these 2 F-MUTs.

In general, data relation $DR$: $Action \times Action \times V \mapsto \{RR, RW, WR, WW, null\}$ between $a_i$ and $a_j$, where $i < j$ can be of 4 kinds:

$$DR(a_i, a_j, v) = \begin{cases} RR & \text{if } \{(a_i, v, r), (a_j, v, r)\} \subseteq Act \\ RW & \text{if } \{(a_i, v, r), (a_j, v, w)\} \subseteq Act \\ WR & \text{if } \{(a_i, v, w), (a_j, v, r)\} \subseteq Act \\ WW & \text{if } \{(a_i, v, w), (a_j, v, w)\} \subseteq Act \\ null & \text{if there is no shared variable.} \end{cases}$$

Accordingly, data dependency $DD$:$Action \times Action \mapsto \{true, false\}$ of actions $a_i$ and $a_j$, where $i < j$, is true if the former action defines (write) a variable that is used (read) by the latter action without any other intermediate definitions:

$$DD(a_i, a_j) = (\exists v \in V \text{ s.t. } DR(a_i, a_j, v)$$
$$= WR \text{ and } \nexists k > i \text{ s.t. } DR(a_k, a_j, v) = WR)$$

Synthesizing a meaningful example is not possible unless each action participating in an example is accompanied by all other actions on which that action is control/data dependent during its execution. We introduce *CDSet*: $Action \mapsto 2^{Action}$, a function that returns the set of actions on which a given action is control dependent, and *DDSet*: $Action \mapsto 2^{Action}$ a function that returns the set of actions on which a given action is data dependent.

Algorithm 2 synthesizes usage examples of the most representative APIs within a given test method *TM*. For each focal method *f* it computes the actions on which the method is dependent for its execution and stores them in the *Result* set. It does the same for all the actions included in the set whose dependencies have not yet been checked (*Result\Processed*). This process continues till the example set *Result* is self-contained, i.e., none of the actions in this set have dependency on actions outside the list. Finally, it returns a set *API-U*, which comprises pairs of a focal method, and a complete usage example of that method synthesized from the *tm*.

---

**Algorithm 2** API usage examples synthesis
1: **procedure** SYNTHESIZEEX(*TM*)
2:     *API−U* ← {}
3:     **for** all $f \in FM_{tm}(TM)$ **do**
4:         *Processed* ← {}
5:         *Result* ← {*f*}
6:         **for** all $a \in$ *Result* \ *Processed* **do**
7:             *Processed* ← *Processed* ∪ {*a*}
8:             *Result* ← *Result* ∪ *DDSet*(*a*) ∪ *CDSet*(*a*)
9:         **end for**
10:        *API−U* ← *API−U* ∪ {(*f*, *Result*)}
11:     **end for**
12: **end procedure**

## 3.3 | Prototype implementation

We implemented the proposed approach to synthesize API uses in an Eclipse plugin. Figure 9 presents the plugin's work flow and input/output for each step. Given a Java project, the plugin distinguishes test code from source code and excludes test cases that may not represent correct API usages. In practice, some tests may be in charge of checking a negative behavior like an exceptional condition. We assume that if a method checks a negative behavior, developers indicate that in JUnit using the expected parameter in @Test annotation, or they explicitly call a JUnit fail method right after an exceptional call in the test case. To discern the role of different methods within a test case, we adapt a graph-based object usage model, *Groum*, that provides necessary control and (intra-procedural) data relations between the statements within a test case.[21] We enrich this partial object usage model with field access information that is computed using the previously developed technique described in Section 2. For each method m, a set MOD(m) of class fields that may be modified by m is computed, and a set USE(m) of class fields that may be used by this method. In both cases, the effects of methods transitively called by m are taken into account. We find internally dependent methods on the basis of the common state, i.e., the fields they read or write. We construct a complete model for each unit test by connecting such methods together. Given an F-MUT f in a test method and a complete object usage model for that test, a slice for f constructs an example of that F-MUT use. Particularly, we realized the slice computation by implementing Algorithm 2 that for each F-MUT in a test, traverses the model starting from the node corresponding to the F-MUT of interest and includes in the program slice every node that can reach the F-MUT using data and/or control edges in the model.

## 3.4 | A complete working example

In this section, we present the detailed operation of our approach through a working example. Figure 10 shows a simple unit test case for the Email class in Commons-Email, an API built on top of the Java Mail API for sending an email. The test case sets some fields of an email message and the outgoing mail server and checks whether a header folding works correctly. Figure 11 illustrates 2 examples constructed from this unit test case. We present the results of different phases, summarized in 4 steps, when the approach is applied to this unit test case.

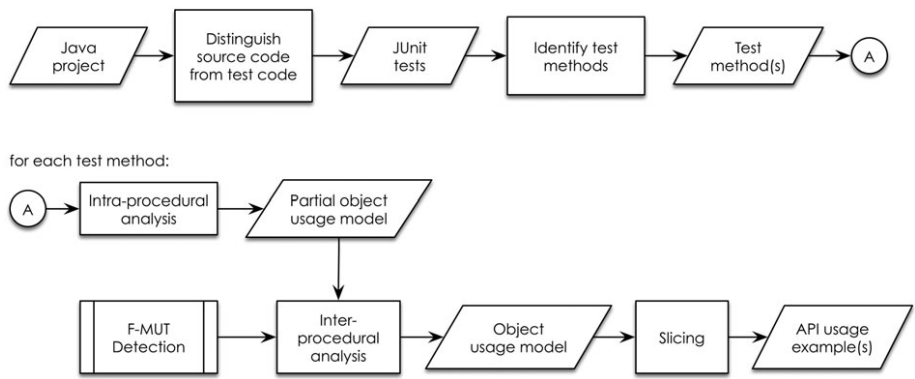

**FIGURE 9** The example synthesis work flow

```
1  public void testFoldingHeaders() throws Exception {
2    email.setHostName(strTestMailServer);
3    email.setSmtpPort(getMailServerPort());
4    email.setFrom("a@b.com");
5    email.addTo("c@d.com");
6    email.setSubject("test mail");
7    final String headerValue = "1234567890 1234567890 123456789 01234567890
        123456789 0123456789 01234567890 01234567890";
8    email.addHeader("X-LongHeader", headerValue);
9    assertTrue(email.getHeaders().size() == 1);
10   // the header should not yet be folded -> will be done by buildMimeMessage
11   assertFalse(email.getHeaders().get("X-LongHeader").contains("\r\n"));
12   email.buildMimeMessage();
13   final MimeMessage msg = email.getMimeMessage();
14   msg.saveChanges();
15   final String[] values = msg.getHeader("X-LongHeader");
16   assertEquals(1, values.length);
17   // the header should be split in two lines
18   final String[] lines = values[0].split("\\r\\n");
19   assertEquals(2, lines.length);
20   // there should only be one line-break
21   assertTrue(values[0].indexOf("\n") == values[0].lastIndexOf("\n"));
22  }
```

**FIGURE 10** A unit test case with dependent sub-scenarios from Commons-Email project

```
 1  final String headerValue = "1234567890 1234567890 123456789 01234567890
        123456789 0123456789 01234567890 01234567890";
 2  email.addHeader("X-LongHeader", headerValue);
```

```
 1  email.setHostName(strTestMailServer);
 2  email.setSmtpPort(getMailServerPort());
 3  email.setFrom("a@b.com");
 4  email.addTo("c@d.com");
 5  email.setSubject("test mail");
 6  final String headerValue = "1234567890 1234567890 123456789 01234567890
        123456789 0123456789 01234567890 01234567890";
 7  email.addHeader("X-LongHeader", headerValue);
 8  email.buildMimeMessage();
```

**FIGURE 11** Synthesized examples from the unit test case in Figure 10

**Step 1. Identifying the analysis scope.** In the first step we identify system classes that are involved in testing. Each system class is analyzed to realize if the class uses any other classes in the system. A call graph is built to represent the calling relationships between the methods of all classes transitively involved in the test.

The approach finds the `Email` class as the only system class in the unit test case. Other classes such as `String` and `MimeMessage` are external classes whose methods have an auxiliary role in the test.

**Step 2. Identifying mutator and inspector methods.** This step tracks the flow of data involving the class fields starting from leaf methods in the call graph and moving toward caller methods used in the test case. It recursively maps field access sets from the callees to the callers in the call graph and categorizes the methods used in the test case into mutators or inspectors according to whether they cause an object's state to change or not, respectively.

Accordingly, `setHostName`, `setSmtpPort`, `setFrom`, `setTo`, `setSubject`, `addHeader`, and `buildMimeMessage` methods are classified as mutators and `getHeaders`, and `getMimeMessage` methods are classified as inspectors.

**Step 3. Identifying F-MUTs.** In each sub-scenario, we identify F-MUT as the last mutator whose effect is inspected in its following assertion expressions.

The approach correctly identifies `email.addHeader` as F-MUT in the first sub-scenario. This method modifies the `headers` field in the `Email` class, and `email.getHeaders`, which is asserted in the first sub-scenario, also reads the same field. Identifying the F-MUT in the second sub-scenario is more interesting. The asserted expression is an array that is initialized by invoking `getHeader` method on an object of type `MimeMessage`. This class is not a system class and therefore cannot be a CUT. In fact, this external object is assigned the result of `getMimeMessage` call on an object of type `Email`, which belongs to the system. Thus, what has been asserted in the second sub-scenario is obtained from the `Email.getMimeMessage`, which inspects the `message` field of the `Email` class and is the actual asserted expression. The last mutator modifying this field is `buildMimeMessage`, which is the F-MUT.

**Step 4. Synthesizing examples.** Our approach relies on data and control dependencies among all expressions in a test to determine which statements are actually necessary to execute each F-MUT.

The approach correctly identifies the declaration of the `headerValue` at line 7 as the only statement in the unit test case on which the focal method `email.addHeader` at line 8 is data dependent. Nonetheless, identifying statements concerning the execution of the second focal method, `email.buildMimeMessage` at line 12 requires an analysis that is finer than the granularity of objects. The approach leverages the field access information, collected in the second step, to perceive the data relationship among the methods. The approach precisely finds that the execution of the `buildMimeMessage` depends on all of its preceding mutator methods.

Finally, an example is constructed on the basis of a set of expression statements on which a given F-MUT is transitively control/data dependent, although in this particular case only the data dependency information is sufficient to construct the example. Figure 11 shows the resulting API usage examples synthesized from the unit test case.

## 4 | EVALUATION

To evaluate this work, we have selected 4 open-source Java projects as the context of our study. Table 3 presents the key characteristics of the subject programs. These are mature programs from different application domains with at least 1 major release. The subject programs are equipped with substantial test suites with test cases in JUnit format. To evaluate our approach we formulated the following research questions:

$RQ_1$: How effective is the proposed approach to identify the relationships between unit tests and source code in the form of F-MUTs?

$RQ_2$: How well can the proposed approach be used to generate meaningful examples from unit test cases?

In the following, we first replicate, on a new machine, an empirical study that we performed in our previous research to assess our approach in identifying F-MUTs ($RQ_1$).[14] We then evaluate whether our example synthesis approach can benefit from the F-MUT information to effectively synthesize meaningful API uses from unit test cases automatically; we compare, on the basis of 3 different metrics, the output of our research prototype with both human-written examples and a state-of-the-art approach (RQ$_2$). The experiments have been conducted on an Intel 2.8 Core i7 CPU machine with 16 GB of RAM and Mac OS X 10.11 operating system.

**TABLE 3** Key characteristics of the subject programs in this study

| | Source code characteristics | | |
|---|---|---|---|
| **Subject Programs** | **Version** | **KLoC** | **Test Methods** |
| Commons Email | 1.3.3 | 8.78 | 130 |
| JGAP | 3.4.4 | 73.96 | 1390 |
| PureMVC | 1.0.8 | 19.46 | 43 |
| XStream | 1.4.4 | 54.93 | 968 |

## 4.1 | Identifying F-MUTs

**Design.** To select test cases for the analysis we first generated a raw dataset by randomly sampling 100 test cases from each of the subject systems. For the `PureMVC` project, which has fewer test cases, we included all its 43 test cases in the dataset. Then, we filtered out test cases that do not satisfy applicability criteria. In particular, we excluded test cases that form an inheritance class hierarchy, do not use standard JUnit assertions, and contain assertion statements in private methods, helper classes, and inherited methods. Most test cases (above 87%) conform to the applicability criteria and test case structure supported by our prototype implementation. Following the preselection and filtering we obtained a dataset of 300 test cases. These test cases form a manually inspected reference dataset (`ref-dataset`$_1$) that we use as an oracle in this study.

For each test case, we have detected F-MUTs manually after thorough analysis of the system specification, javadocs, API usage manuals, and code comments that gave us the knowledge about each system under test. To improve the understanding of concepts between ourselves, we met to identify sub-scenarios and F-MUTs within each test case in 5% of this set. As we assured ourselves that we agreed on how to extract expected information, we (the first 2 authors of the paper) inspected the remaining test cases independently, while providing a short rationale for why each focal method is selected. This rationale is used for internal validation purposes. For all the studied test cases, we exchanged our results to detect potential conflicts. After we agreed on the expected results, we finalized them in the `ref-dataset`$_1$.

To evaluate the effectiveness of the approach, we applied our prototype to all the subject programs and the selected test cases in the `ref-dataset`$_1$. We compared the results of the prototype with the manually identified F-MUTs.

**Results.** The first part of this study involves assessing the effectiveness of the proposed approach to identify F-MUTs; we calculate the precision and recall of our approach with respect to the results in our reference dataset. *Precision* is the fraction of correctly identified F-MUTs versus all results returned by our prototype. *Recall* is the proportion of the results in the `ref-dataset`$_1$ identified by the prototype. To take both into account, we assess the overall effectiveness of our approach using the harmonic-mean ($F_{\beta=1}$) of precision and recall. Table 4 summarizes the results, which confirm our previous findings. Our approach automatically identifies F-MUTs, in a minute in total, and achieves a high precision and a good recall. The harmonic-mean over the 4 subject programs is also promising (66%–87%), which indicates that we can establish the traceability links between unit tests and source code in the form of F-MUTs in an automated manner with a high effectiveness (RQ$_1$). The interested reader may consult our previous work for a detailed discussion on corner cases, as well as some cases where the approach is more advantageous.[14]

**TABLE 4** Quantitative report on the accuracy of the proposed approach on the subject programs

| Subject Programs | Experimental results | | |
| --- | --- | --- | --- |
| | Precision | Recall | H-mean |
| Commons Email | 0.94 | 0.69 | 0.79 |
| JGAP | 0.85 | 0.73 | 0.78 |
| PureMVC | 0.97 | 0.79 | 0.87 |
| XStream | 0.90 | 0.53 | 0.66 |

## 4.2 | Synthesizing API uses

**Design.** For synthesizing API uses we target the same open-source projects that we have used to evaluate identifying F-MUTs (see Table 3). These projects have an established code base and are equipped with documentation. Importantly, we acquired a good knowledge of their APIs during the earlier study and we conducted to assess our approach for identifying F-MUTs.

For evaluating the API use synthesis from test cases, we generated a *raw dataset* comprising 140 test cases, by randomly sampling 35 different test cases from each of the subject systems. We left out simple test cases that only exercise inspector methods within single assertion statements.

To build a reference dataset of API usage examples to be used as an oracle in this study, we recruited 5 Java programmers. All of the programmers had a minimum of 4 years of general programming experience, and on average 2 years expertise in software development with Java programming language. We assigned each participant a random selection of test methods from the raw dataset. We asked the participants to inspect each test method manually and extract from a test any number of meaningful usage examples that it may represent. Each test method was inspected at least by 2 participants; when the participants extracted different usage scenarios from a unit test case, they had to discuss and converge on a set of agreed usage examples. The final results establish the `ref-dataset`$_2$, which consists of 159 human-written examples of API uses.

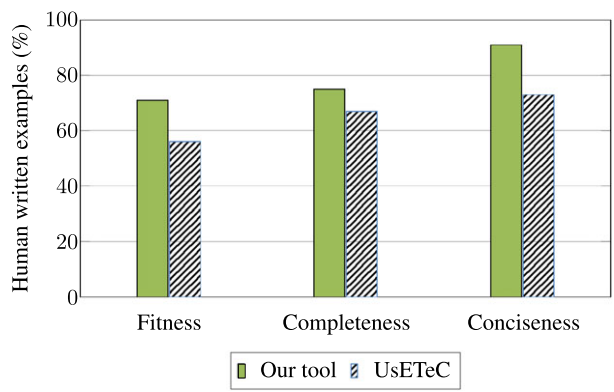

**FIGURE 12** Comparison of the characterization of examples generated by our tool and UsETeC with respect to manually built usage examples

To investigate how well the proposed approach and our prototype implementation that can be used to generate meaningful examples from unit test cases (RQ$_2$), we applied our research prototype and a state-of-the-art tool (UsETeC[4]) to all the subject programs and the selected test cases and compared the outputs to manually built examples in the ref-dataset$_2$.

We evaluate the effectiveness of our approach in comparison with UsETeC, which extracts representative APIs uses from test code.[4] The UsETeC compares the text similarity between a test method name and invoked APIs within the test to find representative APIs. It then uses a slicing technique on the basis of 4 predefined code patterns to extract some code snippets representing usage of these APIs. The code snippets are then clustered to exclude redundant examples.

**Results.** Figure 12 presents the evaluation results and compares the generated examples with the human-written examples in the ref-dataset$_2$. The comparison criteria are *fitness*, *completeness*, and *conciseness*, which we define accordingly.

> **Fitness:** Each test case uses several APIs, including those under test, and others that are used as helpers for initialization or to specify test oracles. We have identified that tested APIs are the most representative ones within a test and consequently a test illustrates more meaningful usage of these APIs (focal use) rather than other non-focal ones. We thus define *Fitness* as the proportion of examples correctly mapped to the focal uses.

The examples produced by our tool are linked to APIs on the basis of F-MUTs; 71% of these examples correspond to the focal uses that are manually picked by a human. In the case of UsETeC, it constructs such links correctly in 56% of cases. The UsETeC relies on name matching to identify focal uses in each test method, which is a brittle technique that only applies to test cases that strictly follow the naming convention.[22-24] In particular, this strategy falls short if the test name does not contain the name of the unit under test or does not entail a known type.

We illustrate the noted shortcoming of UsETeC on the test case in Figure 13. This test case belongs to the *commons email* library and demonstrates how to construct an MIME style email message to send. It sets various email fields and invokes `buildMimeMessage` on the `email` object to actually build the MimeMessage. To examine whether the message is built successfully the unit test invokes `getMimeMessage` at line 10 to get the internal MimeMessage of the `email` object and assigns it to a new MimeMessage object. Invocation of `saveChanges` on this object updates the appropriate header fields of the message to be consistent with the message content. Finally, the test invokes the `getContentType` to retrieve the header field of this message and evaluates via an assertion statement if this is equal to the expected value.

Among several API uses in this unit test case, `buildMimeMessage` is the focal use. In fact, `msg.getContentType()`, the asserted expression in the unit test case invokes the `msg` object, which belongs to `MimeMessage` class, a standard Java library. The test initializes this helper object at line 10 by assigning the `email.getMimeMessage()` method to this object. This invocation is an inspector method that returns the field `Email.message`. Therefore, this method invocation is the actual asserted expression involving the class of the system under test. The method `email.buildMimeMessage()` is the last mutator, which modifies the `Email.message` field accessed in the assertion statement. This method is the F-MUT and also the focal use in this unit test case. Nonetheless, the approach on the basis of name matching cannot identify this focal use correctly using the test name "testDefaultCharsetAppliesToTextContent," as the test name does not entail a known API.

In 71% of cases our approach picks correct focal uses. To further investigate the reasons for missing 29% of correct APIs, we manually inspected test cases corresponding to the missing APIs. We realized that most of these exceptions are due to implementation problems that are reported in Section 2.2, but not the limitations of the approach. Indeed, we could detect focal uses in most of test cases correctly by applying the approach of identifying F-MUTs to each test case manually.

Consider a test case in Figure 14. An external call `iter.remove()` at line 8 affects the oracle part of this test case. An invocation of `map.put()` at line 4 is the last mutator, but the test case focuses on deletion of an item through iteration. In fact, the `iter` object at line 5 holds a reference to the entries in `XmlMap`. That is, the invocation of `remove()` method on the `iter` object at line 8 also removes an item from `XmlMap`. Our prototype does not consider such an implicit dependency between test code and external classes and negatively reports `XmlMap.put()` as the focal use in this test case.

```
1   public void testDefaultCharsetAppliesToTextContent() {
2     email.setHostName(strTestMailServer);
3     email.setSmtpPort(getMailServerPort());
4     email.setFrom("a@b.com");
5     email.addTo("c@d.com");
6     email.setSubject("test mail");
7     email.setCharset("ISO-8859-1");
8     email.setContent("test content", "text/plain");
9     email.buildMimeMessage();
10    final MimeMessage msg = email.getMimeMessage();
11    msg.saveChanges();
12    assertEquals("text/plain; charset=ISO-8859-1", msg.getContentType());
13  }
```

**FIGURE 13** Unit test case for the `Email` class

```
1  public void testRemovesAnItemThroughIteration() {
2    XmlMap map = new XmlMap(this.strategy);
3    map.put("guilherme", "aCuteString");
4    map.put("silveira", "anotherCuteString");
5    for (Iterator iter = map.entrySet().iterator(); iter.hasNext();) {
6      Map.Entry entry = (Map.Entry) iter.next();
7      if (entry.getKey().equals("guilherme")) {
8        iter.remove();
9      }
10   }
11   assertFalse(map.containsKey("guilherme"));
12 }
```

**FIGURE 14** A test method with side effect from an external object

> **Completeness.** Usage examples should be self-contained and include all statements that are necessary for exercising focal uses. Indeed, an example is complete if it includes the appropriate API and all parameters used in the API are well explained (i.e., they are syntactically correct and there is no missing statement) in the code example. We define *completeness* as the proportion of examples that are complete.

Our tool is able to generate about 75% of complete examples, while UsETeC is able to generate 67% of complete examples. The reason that none of the tools is as good as a human at constructing complete examples is due to some dependencies in test methods. For instance, a common practice to decrease code duplication and maintenance costs in testing is to encapsulate the setup part of a test in a helper method that can be called by a group of test methods that need the same test preparation.

In human-written examples, test dependencies are often consolidated into examples. Our tool detects such dependencies, but does not put related code directly into synthesized examples. This is mainly why 25% of examples that our tool generates are marked as incomplete and do not compile. The UsETeC produces even more incomplete examples (33%). This tool suffers from not only the same limitation for test dependencies but also the slicing heuristic, which it applies to test code, which is based on predefined code patterns that in 8% of cases, which was not able to identify all relevant statements to a focal use. We speculate that the slight improvement by our tool is mainly because of the more precise implementation of intra-procedural analysis in our extension of Groum.[21]

> **Conciseness.** To improve readability and enhance developer understanding, usage examples should be concise and free from superfluous statements. In particular, an example is concise if it is complete and does not have more statements than its corresponding human-written example. We define *Conciseness* as the proportion of complete examples that are concise.

Automatically generated examples are less concise than human-written examples. Apart from statements that are substituted for test dependencies in human-written examples, the conciseness of the results is 91% for our tool and 73% for the UsETeC tool (both applied to the complete examples). Human-written examples are more concise mainly because of some code transformation applied during example construction. For instance, the unit test case in Figure 15 checks whether it can properly assign a URL to an EmailAttachment object. The code snippet in Figure 16 illustrates a human-written example corresponding to this test case. It correctly summarizes the intended focal use and also removes unnecessary for statements to demonstrate this usage.

Differently, the slice computation we apply to test code merely removes part of a test that can be found to have no effect on the semantics of interests without any code transformation. Unfortunately, not all examples constructed in this way are sufficiently clear. For example, consider a simple unit test case from the JGap project in Figure 17, and an example corresponding to the third sub-scenario that our tool generates for this unit test case, shown in Figure 18. During slice computation the variable declaration statement in the first sub-scenario is found to be relevant because the chrom object is further used in the third sub-scenario. However, the object definition at line 11 kills the former definition at line 2. That is, the object initialization at the first line of the example is useless and only obscures the focal use. A code transformation that consolidates these statements would make the example far clearer.

In 18% of cases UsETeC generates longer examples than our tool. For instance, Figure 19 shows an excerpt of a unit test case from Commons-Email that we need for our quick discussion. Intuitively, email.addHeader is the focal use in this unit test case, and an example to represent this API use should only include the headerValue declaration statement.

However, UsETeC applied to this unit test case generates a completely different example. Indeed, the focal use addHeader in this test relies on 2 variables, email and headerValue. The UsETeC slices test code on the basis of email, the shared variable among the focal use and other statements, and includes in the same slice all other preceding invocations on this variable that are absolutely irrelevant to illustrate the correct use of addHeader.

The UsETeC does not eliminate statements that may be irrelevant in an example, e.g., external APIs, and inspector methods. It is confined to the intra-procedural relations and roughly assumes 2 statements are dependent and should participate in an example if they involve the same object(s), which is not true in general, as discussed in Section 3. We attribute our 18% improvement to more precise test analysis, i.e., realizing inter-procedural data relations (13%) and control dependencies (5%) among statements.

```
1  public void testGetSetURL() throws Exception {
2    final String[] tests =
3    {
4      "http://localhost/",
5      "http://www.apache.org/",
6      "http://foo.notexisting.org" };
7
8    for (final String urlString : tests)
9      {
10        final URL testURL = new URL(urlString);
11        attachment.setURL(testURL);
12        assertEquals(testURL, attachment.getURL());
13     }
14 }
```

**FIGURE 15** A simple unit test case

```
1    String urlString = "http://localhost/";
2    EmailAttachment attachment = new EmailAttachment();
3    final URL testURL = new URL(urlString);
4    attachment.setURL(testURL);
```

**FIGURE 16** A human-written example corresponding to the unit test case in Figure 15

```
1  public void testConstruct_3() throws Exception {
2    Chromosome chrom = new Chromosome(conf, new IntegerGene(conf), 1);
3    assertEquals(1, chrom.size());
4
5    chrom = new Chromosome(conf, new IntegerGene(conf), 1, null);
6    assertEquals(1, chrom.size());
7    assertNull(chrom.getConstraintChecker());
8
9    /* MetaComment: third sub-scenario starts here */
10   IGeneConstraintChecker cc = new MyConstraintChecker();
11   chrom = new Chromosome(conf, new IntegerGene(conf), 1, cc);
12   assertEquals(1, chrom.size());
13   assertEquals(cc, chrom.getConstraintChecker());
14 }
```

**FIGURE 17** Unit test case for the `Chromosome` class

```
1  // what our tool generates
2  Chromosome chrom = new Chromosome(conf, new IntegerGene(conf), 1);
3  IGeneConstraintChecker cc = new MyConstraintChecker();
4  chrom = new Chromosome(conf, new IntegerGene(conf), 1, cc);
```

```
1  // what human generates after transformation
2  IGeneConstraintChecker cc = new MyConstraintChecker();
3  Chromosome chrom = new Chromosome(conf, new IntegerGene(conf), 1, cc);
```

**FIGURE 18** An example representing the use of the `Chromosome` class in the third sub-scenario in Figure 17

## 4.3 | Threats to validity

We note several limitations and threats to validity of the assessment of our approach. We mitigated the risks to the external validity and the generalization of the results by selecting real-world systems from different application domains with manually generated test cases. The study involved a randomly selected sample of test cases from the different systems.

Threats to internal validity might arise from the process used in our empirical study. We used statistical methods to evaluate the result of the experiments where results could have been affected by a randomness in the test case selection. The accuracy of the results used to evaluate our approach affect the results achieved. We did not have access to the original program developers to indicate the focal methods in each test case. For this reason, we familiarized ourselves with the project documentation and details of the source and test code and

```
 1  public void testFoldingHeaders() throws Exception {
 2    email.setHostName(strTestMailServer);
 3    email.setSmtpPort(getMailServerPort());
 4    email.setFrom("a@b.com");
 5    email.addTo("c@d.com");
 6    email.setSubject("test mail");
 7    final String headerValue = "1234567890 1234567890 123456789 01234567890
         123456789 0123456789 01234567890 01234567890";
 8    email.addHeader("X-LongHeader", headerValue);
 9    assertTrue(email.getHeaders().size() == 1);
10    assertFalse(email.getHeaders().get("X-LongHeader").contains("\r\n"));
11    /* MetaComment: the rest of code is omitted to save space */
12  }
```

**FIGURE 19**   The first sub-scenario of the unit test case shown in Figure 10

cross-checked the results. We acknowledge that the approach validated against manual analysis that has been performed by 2 of the authors is a threat to validity because of potential bias.

Furthermore, to mitigate the construct validity risks and avoid mono-method bias in our experiments, we use 3 complementary metrics (fitness, completeness, and conciseness) for generalizing the quality and meaningfulness of the generated API usage examples. Nonetheless, studying these metrics may not reflect all the properties of the generated usage examples.

Our approach shares inherent limitations with other static analysis techniques that are generally not sound. According to our findings in this study, many of the false negatives are due to implementation problems reported in Section 2.2, but not the limitations of the approach. Nevertheless, static analysis is preferable for analyzing large code bases. It reduces the cost of the application of our approach when it comes to minimizing the time to setup execution environments, working machines, etc, required to run the programs for dynamic analyses. Using better points-to analysis for type resolution and more precise call graph construction, for instance, using Spark[25] could improve the approach.

Although the aim of this research is to provide developers with API usage examples mined from unit test cases when other sources of client code is inaccessible, we did not evaluate the effectiveness of extracting usage examples from test cases in comparison with those extracted from other sources, such as client systems. This would be important to decide the best source of usage examples when different sources are available. We leave this experiment to future work.

## 5 | RELATED WORK

This section overviews the related work concerning the 2 necessary phases of the approach, i.e., to test code traceability link recovery and example extraction.

### 5.1 | Linking test code and source code

The practical automated realization of test-to-code traceability has received little attention. Van Rompaey and Demeyer compare several traceability resolution strategies to link test cases and the units under test.[23] For example, in the naming convention strategy, they match production code and test code by removing the string "test" from the name of the test case. This strategy falls short if the test name does not contain the name of the unit under test or does not entail a known type. In another strategy, they analyze call behavior before assertion statements and presume that a test case calls a method on the unit under test right before the assertion statement. They exploit the static call graph to identify the last class called before an assert statement. This strategy fails when, right before the assert statement, there is a call to a class other than the tested class.[24]

A strategy on the basis of lexical analysis builds on an assumption that developers use similar vocabulary to write the source code of a test case and the corresponding unit under test. Latent semantic indexing, an information retrieval technique, is used to calculate this similarity. However, their study shows that a significant amount of vocabulary in a test case does not repeat in the unit under test. Finally, a version log mining strategy builds on an assumption that test cases and their corresponding unit under test coevolve together throughout time. This strategy bears a risk to wrongly identify production code that changes frequently as the unit under test.

Qusef et al. propose to use data flow analysis to circumvent the limitations of these strategies.[24] They apply reachability analysis and exploit data dependence to identify a set of classes that affect the result of the last assertion statement in each unit test. This analysis, however, does not consider inter-procedural flow, inheritance, and aliasing. The SCOTCH is an improvement over this work—a technique on the basis of dynamic slicing to restore test case traceability links.[26] The set of identified classes by dynamic slicing is an overestimate of the set of classes actually tested in a test case. In fact, a slice will contain all the helper classes used in a test case as well. In a recent work, the same authors use another filtering strategy on the basis of name similarity to enhance the accuracy of their earlier approach.[27] These approaches rely on a "stop-class list" to hold the names of the classes to be considered as helper class in the analysis; however, these classes have to be manually identified prior to the analysis.

While these works realize test to source code traceability links on the class level, Gälli et al. provide initial evidence that a single method is most often the unit under test in object-oriented programs.[17] Nevertheless, there is a scant work on automatically identifying methods under test. Marschall exploits the naming convention of a test method to establish the relation between tests and methods under test.[22] This requires developers to strictly follow the naming conventions, which is not regular in practice.[24] Additionally, a test case may have different sub-scenarios and multiple methods under test, accordingly. But the test name may not entail the information about all these methods.

To improve this approach, Marschall also suggests a method as a tested method, if it creates an argument for an assertion statement. This produces many false positives as it reports all the inspector methods whose results are asserted in a test case. Ying et al. propose a call graph filtering approach to detect methods that are probably irrelevant during program investigation.[28] According to their findings, methods closer to the leaf of a call graph, as well as those with a small number of callees are unlikely to contribute to the understanding of the application logic. They use this approach to eliminate irrelevant methods from the set of methods that can be invoked, transitively, from a JUnit test case. This heuristic highlights the setup parts of a test and misses to detect a tested method, which is called right before an assertion. In addition, it fails to retrieve relevant invocations in a test with multiple sub-scenarios. In a recent work, Ghafari et al. propose an approach to automatically establish the relationship between the source and test code on the method level.[14] They use classic analysis techniques to dissect the structure of unit test cases and realize the role of each method within a unit test. As a result, they precisely detect the focal methods that represent the core of a test scenario inside a unit test case. This precursor work, indeed, has inspired and enabled the presented approach in this paper.

## 5.2 | Example extraction

Extensive research has been performed to mine API usages from a local repository. Mandelin et al. observe that usually a programmer knows what type of object she needs, but does not know how to write the code to get the object.[29] To enable code reuse, they develop techniques for synthesizing code fragments automatically given a simple query that describes that desired code of input and output types. Zhong et al. propose one of the first approaches that mine common API usage patterns for suggesting relevant code snippets to aid developers.[3] Lack of code examples in API documentation motivates Kim et al. to propose a code example recommendation system that augments the standard API documentation with the code examples organized locally.[30] Buse et al. present a technique to automatically synthesize human-readable API usage examples from a given software corpus to enrich documentation.[5] This approach is the first to leverage type information and statement ordering; and generated examples are free from superfluous context and contain only the program statements needed to show typical use of a target data type. Likewise, Montandon et al. instrument API documentation in JavaDoc format with concrete usage examples.[12] Wang et al. propose a technique to improve the quality of usage patterns.[31] This work proposes a 2-step clustering algorithm to produce succinct and high-covering examples. While the mentioned approaches usually generate small examples, which are helpful in the initial stage of API learning, Moritz et al. propose a visualization-based approach for finding more detailed API usage examples that are helpful in later stages.[32] Galenson et al. propose an approach for code synthesis given only partial specifications of desired results.[33] In contrast to most of the approaches that rely on static information, this approach is dynamic and also allows users to incrementally give more information to refine the candidate code fragments. Nguyen et al. propose an API recommendation approach premised on statistical learning from low-level code changes and the context of such changes.[34] It suggests a desirable API call in the top 5 positions in 77% of the time, which is a significant improvement over state-of-the-art approaches.

The huge amount of open-source code that is available online has motivated researchers to focus on realizing internet-scale code search to enable developers to reap the benefits of these billions of lines of source code. However, ranking high-quality code examples at the top of the result set is challenging. Chatterjee et al. provide a code search technique that locates a small set of relevant code snippets to perform a desired task specified with a free-form query.[35] To improve search results, this technique relies on the API documentation to annotate undocumented code with the plain English meaning of each API. Although this approach provides better results than existing code search engines, its time complexity is high at runtime, which decreases its usability for real applications with a large-scale corpus. To address this issue, Keivanloo et al. propose an approach to answer a free-form query within hundreds of milliseconds on a corpus covering millions of code snippets.[6] This approach enables working code examples to be spotted in a time complexity similar to internet-scale code search engines. Existing code search engines use ranking algorithms that only compute the relevancy between the query and the results. Consequently, they may produce some results that all share the same characteristics, e.g., the code snippets originate from the same project with the same implementation. Hence, other different results with lower relevance scores on the basis of the query are hidden, even though they may actually be closer to what a programmer is looking for. Martie et al. propose 4 ranking algorithms that take into account relevance, diversity, and conciseness in ranking code search results.[36] Two of these algorithms, which leverage the social and technical information of the code results, produce top 10 results that are much preferable to use by the programmers.

Recent research also relies on communities where developers post questions and receive answers regarding their programming issues to extract useful development hints. These crowd sources often become an alternative for official API documentation where such documentation is either sparse or unavailable. De Souza et al. present an approach that leverages the Stack Overflow knowledge (questions and answers) to recommend information that is useful for developers.[37] They present this information in a web browser and rank the results on the basis of the textual similarity between the query and the pairs of questions and answers, as well as the quality of the pairs. However, this approach comes with a number of problems. Every time developers need to look for information, they interrupt their work flow and switch from the IDE to a web browser to perform and refine searches. Besides, query formulation is not easy to accomplish manually. Developers may not know what to search for and how to formulate their needs in a query appropriately. Ponzanelli et al. propose an approach to retrieve pertinent Stack Overflow discussions to the code context in the IDE.[9] For each retrieved discussion, they calculate textual and structural similarity of the code context and the discussion, as well as the rate of that discussion on Stack Overflow to rank possible solutions.

While all the aforementioned approaches rely on the client code, Ghafari et al. propose the idea of extracting the API usage examples from unit test cases.[11] Zhu et al. present the UsETeC tool to extract such examples from test code.[4] However, to find representative APIs within a test case they rely on a test method naming convention that has been proven to be inapplicable for this purpose.[23,24] Conversely, we identify F-MUTs that represent the actual intents of a unit test case.[14] In addition, UsETeC applies a heuristic code slicing technique that is restricted to predefined code patterns without realizing the exact relationship among code elements and their roles within a test, whereas we perform a more precise test analysis to perceive the role of different API calls within a test and synthesize examples that are highly similar to human generated ones.

## 6 | CONCLUSION

Developers cope with the complexity of modern software systems and strive to speed up the development process by increasingly relying on the functionalities provided by off-the-shelf components and frameworks. However, understanding how to properly use APIs of large libraries is non-trivial and requires extensive developer learning and effort. With the rise of the open-source movement, an increasing quantity of source code is becoming available in public, and software developers are resorting to study and reuse existing source code to alleviate the aforementioned problems. However, manually finding sufficient and adequate code examples for a given API that is not necessarily designed in a reusable fashion is a difficult and time-consuming activity.

In this paper we proposed an approach that only depends on unit test cases of an API to generate usage examples. We implemented the proposed approach in a research prototype that applies to object-oriented software and unit test cases in JUnit format. The approach requires API source code and its unit test cases and uses classic static analysis techniques to extract the information required to build the API usage examples. While being superior to the state-of-the-art, the proposed approach provides API usage examples that are highly similar to human-written examples and are particularly helpful to both the API developers and the API clients in the absence of client code.

### ACKNOWLEDGMENTS

We appreciate the valuable contribution of Prof. Carlo Ghezzi, the first author's advisor during his PhD studies at Politecnico di Milano. We also gratefully acknowledge the financial support of the Swiss National Science Foundation for the project "Agile Software Analysis" (SNSF project No. 200020-162352, January 1, 2016-December 30, 2018).